\documentclass[prd,twocolumn,showpacs,floatfix,superscriptaddress]{revtex4}
\usepackage{graphicx}
\usepackage{epsfig}
\usepackage{bm}
\usepackage{amssymb}
\usepackage{float}
\usepackage{amsmath}
\usepackage{dcolumn}
\usepackage{cancel}
\usepackage[colorlinks]{hyperref}
\usepackage[usenames,dvipsnames]{color}
\hypersetup{
     breaklinks=true,
    pdfstartview={FitH},    
    colorlinks=true,       
    linkcolor=blue,          
    citecolor=red,        
    filecolor=magenta,      
    urlcolor=blue,           
    anchorcolor=green,      
    linktocpage=true
}

\def\doi{http://doi.org}
\def\arxiv{http://arxiv.org/abs}

\begin{document}

\title{Ricci-Gauss-Bonnet holographic dark energy}

\author{Emmanuel N. Saridakis}
\email{Emmanuel\_Saridakis@baylor.edu}
\affiliation{Chongqing University of Posts \& Telecommunications, Chongqing, 400065, 
China}
\affiliation{Department of Physics, National Technical University of Athens, Zografou
Campus GR 157 73, Athens, Greece}

\affiliation{CASPER, Physics Department, Baylor University, Waco, TX 76798-7310, USA}

\begin{abstract} 
We present a model of holographic dark energy in which the Infrared cutoff 
is determined by both the Ricci and the Gauss-Bonnet invariants. Such a 
construction has the significant advantage that the Infrared cutoff, and consequently the 
holographic dark energy density, does not depend on the future or the past evolution of 
the universe, but only on its current features, and moreover it is determined by 
invariants, whose role is fundamental in gravitational theories. We extract 
analytical solutions for the behavior of   the dark energy density and equation-of-state 
parameters as functions of the redshift. These reveal the usual thermal history 
of the universe, with the sequence of radiation, matter and dark energy epochs, resulting 
in the future to a complete dark energy domination. The corresponding dark energy 
equation-of-state parameter can lie in the quintessence or  phantom regime, or 
experience the phantom-divide crossing during the cosmological 
evolution, and its asymptotic value  can be quintessence-like, phantom-like, or be 
exactly equal to the cosmological-constant value. Finally, we extract the 
constraints on the model parameters that arise from Big Bang Nucleosynthesis. 
\end{abstract}


\maketitle

\section{Introduction}

According to the concordance  paradigm of cosmology the universe, after being in a long 
matter-dominated epoch, has entered in an accelerating phase. Although the reasonable 
explanation is that the cause of this acceleration is the cosmological constant 
\cite{Peebles:2002gy}, the possibility of a dynamical nature, as well as the need to 
additionally describe the early-time acceleration era (inflation) too, led to the need of 
introducing extra degree(s) of freedom beyond the standard framework of general 
relativity and Standard Model of particles. One can attribute these extra degrees of 
freedom to new, exotic forms of matter, such as in  the dark energy concept  
(for reviews see \cite{Copeland:2006wr,Cai:2009zp}), or alternatively, one can consider 
the extra degrees of freedom to have gravitational origin, i.e. to arise from a 
gravitational modification that possesses general relativity as a particular limit (see 
\cite{Nojiri:2006ri,Capozziello:2011et,Cai:2015emx}).

A rather interesting alternative for the explanation of the dark energy nature is 
obtained through application of the fundamental holographic principle, that 
arises from black hole thermodynamics and string theory 
\cite{tHooft:1993dmi,Susskind:1994vu,Witten:1998qj,Bousso:2002ju},
at a cosmological framework \cite{Fischler:1998st,Bak:1999hd,Horava:2000tb}. Starting from 
the connection 
of the Ultraviolet cutoff of a quantum field theory, which is related to the vacuum 
energy, with the largest distance of this theory,  which is necessary for the  quantum 
field theory  applicability at large distances \cite{Cohen:1998zx}, then the obtained 
vacuum energy will be a form of dark energy of holographic origin, called holographic 
dark energy \cite{Li:2004rb} (for a review see \cite{Wang:2016och}). Despite some 
objections 
\cite{Easther:1999gk,Kaloper:1999tt}, holographic dark energy has attracted a considerable 
amount of research and 
proves to have interesting cosmological phenomenology either in its basic 
\cite{Li:2004rb,Horvat:2004vn,Huang:2004ai,Pavon:2005yx,Wang:2005jx,Nojiri:2005pu,
Kim:2005at,
Wang:2005ph, Setare:2006wh,Setare:2008pc,Setare:2008hm} as well 
as in its various extensions 
\cite{Gong:2004fq,Saridakis:2007cy,  
Setare:2007we,Setare:2008bb,Saridakis:2007ns,Saridakis:2007wx,Jamil:2009sq,
Gong:2009dc,BouhmadiLopez:2011xi,Malekjani:2012bw,Khurshudyan:2014axa,
Landim:2015hqa,Pasqua:2016wrm,
Jawad:2016tne,Pourhassan:2017cba}, while it has been shown to be in agreement with 
observational data 
\cite{Zhang:2005hs,Li:2009bn,Feng:2007wn,Zhang:2009un,Lu:2009iv,Micheletti:2009jy}.

Although the basic idea that holographic dark energy density should be proportional to 
the inverse squared Infrared cutoff $ L$, namely
 \begin{equation}
 \label{basic}
\rho_{DE}=\frac{3c}{\kappa^2 L^2},
\end{equation}
with $\kappa^2$ the gravitational constant and $c$ a parameter, is in general accepted as 
long as one agrees with the cosmological application of holographic principle, there is 
not a concrete answer and consensus on what the Infrared cutoff should be. Since the 
simplest choice, namely the Hubble radius, cannot lead to an accelerating universe 
\cite{Hsu:2004ri}, and since the next guess, namely the   particle horizon, cannot 
drive acceleration either, in the original version of the scenario it was the future 
event horizon that was finally used \cite{Li:2004rb}. Although such a choice leads to 
interesting cosmological implications, it possesses an unpleasant feature, namely that 
the present value of the dark energy density is determined by its future evolution. 
Hence, various ``modified'' models of holographic dark energy appeared in the literature, 
in which the Infrared cutoff is not the future event horizon, but   quantities with 
dimensions of lengths that depend on the past or the present features of the universe. In 
the first class of models one can have the ``agegraphic dark energy'' scenario, in which 
the age of the universe or the conformal time play the role of the Infrared cutoff
 \cite{Cai:2007us,Wei:2007ty,Wei:2007ut,Jamil:2010vr}, while in the second class the 
inverse square root of the Ricci curvature is used 
\cite{Gao:2007ep}. 

The Ricci holographic dark energy, apart from the advantage that it does not depend 
neither on the future nor on the past universe evolution, and apart from its interesting
cosmological applications  
\cite{Gao:2007ep,Feng:2009ai,Feng:2009jr,Suwa:2009gm,Belkacemi:2011zk}, has the advantage 
that the Infrared cutoff is calculated through a gravitational invariant. Although the 
underlying theory that could lead to that is unknown, such a possibility is theoretically 
intriguing since the use of invariants in physics is fundamental. Nevertheless, apart from 
the Ricci scalar, one could have other invariants determining the Infrared cutoff too, and 
the simplest such extension would be to use the Gauss-Bonnet combination. 

In this work we are interested in exploring such a possibility, i.e. to construct a model 
of holographic dark  energy in which the Infrared cutoff is determined by both the Ricci 
scalar and the Gauss-Bonnet invariant. Such a model is theoretically more concrete, since 
higher order invariants contribute too, and it has an additional parameter that allows 
for richer cosmological behavior. The plan of the manuscript is the following. In Section 
\ref{model} we present the scenario of Ricci-Gauss-Bonnet holographic dark energy. In 
Section \ref{Cosmologicalevolution} we investigate the cosmological applications, 
extracting analytical expressions and studying the evolution of dark energy density and 
equation-of-state parameters. Finally, Section \ref{Conclusions} is devoted to the 
conclusions.

\section{Ricci-Gauss-Bonnet holographic dark energy}
\label{model}
  
In this section we will  construct a model of holographic dark energy, in which the 
Infrared cutoff will be a combination of the Ricci and Gauss-Bonnet scales.
We consider a homogeneous and isotropic Friedmann-Robertson-Walker (FRW)
  metric of the form
\begin{equation}
\label{FRWmetric}
ds^2=- dt^2+a^2(t) \left(\frac{dr^2}{1-kr^2}+r^2 d\Omega^2\right),
\end{equation}
where $a(t)$ is the scale factor and with
$k=0,+1,-1$ corresponding to flat, close and open spatial geometry respectively. In the 
following we will focus on the flat case for convenience, however the generalization to 
non-flat geometry is straightforward.   

In usual 
Ricci dark energy \cite{Gao:2007ep} one uses the Ricci scalar calculated in an FRW metric 
in order to account for the IR cutoff. In particular, since the Ricci scalar has 
dimensions of inverse length square, one obtains a holographic dark energy density 
proportional to it. Nevertheless, it is known that when one uses curvature 
invariants in a particular modification, for consistency he should use other invariants 
that could participate in the same order too. Hence, since in FRW  geometry the 
Gauss-Bonnet combination 
$G=R^2-4R_{\mu\nu}R^{\mu\nu}+R_{\mu\nu\rho\sigma}R^{\mu\nu\rho\sigma}$ is of the order of 
$R^2$, in a model of holographic dark energy where $R$ is used to determine the IR 
cutoff, $\sqrt{|G|}$ should be contribute too. Having these in mind, in this work we will 
consider a scenario of holographic dark energy in which the inverse squared IR cutoff is
\begin{equation}\label{GBHDEL}
\frac{1}{L^2}=-\alpha R+\beta \sqrt{|G|},
\end{equation}
where the constants  $\alpha$ and $\beta$ are the model parameters. Clearly, for 
$\beta=0$ on re-obtains the standard Ricci dark energy, while for $\alpha=0$ we acquire a 
pure Gauss-Bonnet holographic dark energy.
Inserting (\ref{GBHDEL}) into (\ref{basic}),  we obtain the energy density of the 
Ricci-Gauss-Bonnet 
holographic dark energy as
\begin{equation}\label{rhoHDE}
 \rho_{DE}=\frac{3}{\kappa^2}\left(-\alpha R+\beta \sqrt{|G|}
 \right),
\end{equation}
where we have absorbed the constant $c$ into $\alpha$ and $\beta$.
In flat FRW geometry the Ricci scalar and the Gauss-Bonnet combination read as
 \begin{eqnarray}
 && R=-6\left(2H^2+\dot{H}\right)
 \label{Ric}\\
 &&G=24H^2\left(H^2+\dot{H}\right)
  \label{GB},
 \end{eqnarray}
where $H=\dot{a}/a$ is the Hubble function, with dots denoting derivatives with respect 
to cosmic time $t$. Therefore, the Ricci-Gauss-Bonnet dark energy density becomes
 \begin{equation}\label{rhoHDE2}
 \rho_{DE}=\frac{3}{\kappa^2}\left[6 \alpha \left(2H^2+\dot{H}\right)+2\sqrt{3} \beta H
\sqrt{|H^2+\dot{H}|}
 \right].
\end{equation}
The first Friedmann equation writes as
\begin{equation}\label{FR1}
H^2=\frac{\kappa^2}{3}\Big(\rho_m+ \rho_{DE} \Big),
\end{equation}
with $\rho_m$ the energy density of the matter sector, which as usual is assumed to 
correspond to a perfect fluid,   whose equation-of-state parameter is $w_m=p_m/\rho_m$ 
with $p_m$ its pressure. The equations close by considering the usual conservation 
equation for the matter sector, namely 
\begin{equation}\label{rhocons}
\dot{\rho}_m+3H(\rho_m+p_m)=0.
\end{equation}

Equations (\ref{FR1}) and (\ref{rhocons}) can determine the universe evolution as long as 
the matter equation of state is known. In particular, considering a pressureless 
(i.e. $w_m=0$) matter sector, equation (\ref{rhocons}) gives $\rho_m=\rho_{m0}/a^3$, with 
$\rho_{m0}$ the value of the matter energy density at the present scale factor $a_0=1$ 
(in the following the subscript ``0'' marks the value of a quantity at present). Thus, 
inserting into (\ref{FR1}), and knowing  (\ref{rhoHDE2}), we obtain a differential 
equation 
for $a(t)$ that can be solved similarly to all dark-energy and modified-gravity models. 
Nevertheless, in the following we desire to investigate the scenario at hand elaborating 
the equations suitably, and provide analytical expressions.

\section{Cosmological evolution}
\label{Cosmologicalevolution}

Let us elaborate the equations in a similar way to many holographic dark energy models, 
and provide analytical solutions, focusing for simplicity to the usual dust matter case. 
It proves convenient to introduce the density parameters as 
 \begin{eqnarray}
 && \Omega_m\equiv\frac{\kappa^2}{3H^2}\rho_m
 \label{Omm}\\
 &&\Omega_{DE}\equiv\frac{\kappa^2}{3H^2}\rho_{DE}
  \label{OmDE},
 \end{eqnarray}
 which for dust matter gives immediately $\Omega_m=\Omega_{m0} H_0^2/(a^3 H^2)$. Knowing 
that in terms of the density parameters  the Friedmann equation (\ref{FR1}) becomes 
just $\Omega_m+\Omega_{DE}=1$, we can easily extract that  
 \begin{equation}\label{Hrel}
H=\frac{H_0\sqrt{\Omega_{m0}}}{\sqrt{a^3(1-\Omega_{DE})}}.
\end{equation}
As usual, instead of time $t$ it proves more convenient to use $x\equiv \ln a$ as the 
independent variable, and hence for every quantity $f$ we will have $\dot{f}=f' H$, with 
primes denoting derivatives in terms of $x$. Differentiating (\ref{Hrel}) we find 
  \begin{equation}\label{Hdorrel}
\dot{H}=-\frac{H^2}{2(1-\Omega_{DE})}\left[3(1-\Omega_{DE})-\Omega_{DE}'
\right].
\end{equation}
 Hence, inserting (\ref{Hdorrel}) into (\ref{Ric}) and (\ref{GB}) we result to 
  \begin{eqnarray}
 && R=-3 H^2 \left[1+\frac{\Omega_{DE}'}{1-\Omega_{DE}}
 \right]
 \label{RicH2}\\
 &&G=12H^4 \left[\frac{\Omega_{DE}'}{1-\Omega_{DE}}-1
 \right]
  \label{GBH2} .
 \end{eqnarray}
 Finally, inserting these into (\ref{rhoHDE}) and then into (\ref{OmDE}) we   
obtain 
   \begin{equation}
  \Omega_{DE}=3\alpha \left[1+\frac{\Omega_{DE}'}{1-\Omega_{DE}}
 \right]+2\sqrt{3}\beta \sqrt{\left|\frac{\Omega_{DE}'}{1-\Omega_{DE}}-1
 \right|}
  \label{OmDEdifeq} .
 \end{equation}
 This is the differential equation that determines the evolution of Ricci-Gauss-Bonnet 
holographic dark energy, in a flat universe and for dust matter. This 
equation accepts an analytic solution in an implicit form, namely:
   \begin{eqnarray}
 &&  \!\!\!\!\!\!\!
 \epsilon_{\pm}  \frac{\gamma_+}{\delta \zeta_+}\, \text{arctanh} 
\left(-\frac{\sqrt{6\alpha^2+\beta^2-\alpha\,\Omega_{DE}}}{ \zeta_+}
 \right)\nonumber\\
  && \!\!\!\! \! \!\!  -  \epsilon_{\pm}   \frac{\gamma_-}{\delta \zeta_-}\, \arctan 
\left(-\frac{\sqrt{6\alpha^2+\beta^2-\alpha\,\Omega_{DE}}}{ \zeta_-}
 \right)\nonumber\\
 &&\!\!\!\!\!\!\!
 - 24 \epsilon_{\pm} 
\frac{\beta}{\delta}\sqrt{6\alpha^2+\beta^2-\alpha}    
 \, \text{arctanh} 
\left(\!\frac{\sqrt{6\alpha^2\!+\!\beta^2\!-\!\alpha\,\Omega_{DE}} 
}{\sqrt{6\alpha^2\!+\!\beta^2\!-\!\alpha}}
 \right)
 \nonumber\\
 &&\!\!\!\!\!\!\!
-\frac{\sqrt{3}(3\alpha^2\!+\!2\beta^2\!-\!\alpha)}{\delta}\,\ln\left[\frac{
9\alpha^2\!-\!12\beta^2\!-\!6\alpha\,
\Omega_{DE}\! +\!\Omega_{DE}^2} {(\Omega_{DE}-1)^2 }
 \right]
 \nonumber\\ 
 &&\!\!\!\!\!\!\!
 -\frac{2 (9\alpha-1)\beta}{\delta}
 \, \text{arctanh} 
\left(\frac{\Omega_{DE}-3\alpha}{2\sqrt{3}\beta}
 \right)=2\ln a+x_0,
 \label{gensol}
 \end{eqnarray}
where $\epsilon_{\pm}=\pm1$ determines the two solution branches. In the above 
expression we have defined the constants
 \begin{eqnarray}
 && \!\!\!
 \gamma_{\mp}=2\left[\mp  9 \alpha^3
 +
\beta^2\left(\pm1+2\sqrt{3}\beta\right)
\right.\nonumber\\
&& \ \ \ \ \ \ \ \  \ 
\left. 
 \mp\alpha \beta \left(\pm 
2\sqrt{3}+15\beta\right)
+3\alpha^2 
\left(4\sqrt{3}\beta\pm1\right)
 \right],\nonumber\\
 &&\delta=\frac{1-6\alpha+9\alpha^2-12\beta^2}{\sqrt{3}},\nonumber\\
 &&\zeta_{\mp}=\sqrt{\mp3 \alpha^2+2\sqrt{3}\alpha\beta\mp\beta^2}.
 \end{eqnarray}
 Furthermore, $x_0$ is an integration constant which can be determined applying 
(\ref{gensol}) 
at present, i.e. at $a=a_0=1$, and setting $\Omega_{DE}=\Omega_{DE0}$. In summary, 
(\ref{gensol}) provides the 
two branches of the analytical solution of the Ricci-Gauss-Bonnet holographic dark energy 
as a function of the logarithm  of the scale factor. Hence, one immediately has its 
behavior in terms of the redshift $z=a_0/a-1$, 
since $x\equiv\ln a=-\ln(1+z)$. Lastly, inserting   (\ref{gensol}) into 
(\ref{Hrel}) provides $H(z)$,  and moreover performing the integration one can obtain the 
solution for $a(t)$ too, however this is not necessary since knowing the behavior of the 
observables in terms of $z$ is adequate.

Apart from the evolution of the dark energy density parameter $\Omega_{DE}(z)$ that was 
extracted above, the other important observable is the dark energy equation-of-state 
parameter $w_{DE}$. In order to calculate it we proceed as follows. Since matter is 
conserved according to (\ref{rhocons}), we deduce that as usual holographic dark energy 
is conserved too, namely
\begin{equation}\label{rhoDEcons}
\dot{\rho}_{DE}+3H(1+w_{DE})\rho_{DE}=0,
\end{equation}
where $w_{DE}\equiv 
p_{DE}/\rho_{DE}$, with $p_{DE}$ the holographic dark energy pressure. Inserting 
(\ref{rhoHDE}) into 
(\ref{rhoDEcons}) we obtain
\begin{equation}\label{wDEgen}
 w_{DE}=-1+\Omega_{DE}^{-1}\left[\frac{\alpha R'}{3 H^2}-\frac{\beta 
|G'|}{6H^2\sqrt{|G|}}
 \right],
\end{equation}
where as previously primes denote derivatives with respect to $x\equiv\ln a$. 
Differentiating  
(\ref{Ric}) and (\ref{GB}) and using (\ref{Hdorrel})  we straightforwardly acquire 
 \begin{equation}
   \frac{R'}{3H^2}=3+\frac{2\Omega_{DE}'}{1-\Omega_{DE}}- 
\frac{2(\Omega_{DE}')^2}{(1-\Omega_{DE})^2}-\frac{\Omega_{DE}''}{1-\Omega_{DE}}
\label{Ricprim}
 \end{equation}
and 
 \begin{eqnarray}
 &&   \frac{ 
|G'|}{6H^2\sqrt{|G|}}=\frac{1}{\sqrt{3}(1-\Omega_{DE})^2}
\frac{1}{\sqrt{\left|\frac{\Omega_{DE}'}{1-\Omega_{DE}}-1
\right|}}\nonumber\\
&&
\cdot\Big\{2\left[\Omega_{
DE}'-3(1-\Omega_{DE})
\right](\Omega_{DE}'+\Omega_{DE}-1)\nonumber\\
&&
+\Omega_{DE}''(1-\Omega_{DE})+(\Omega_{DE}
')^2
\Big\}.
\label{GBprim}
 \end{eqnarray}
Thus, inserting (\ref{Ricprim}),(\ref{GBprim}) into (\ref{wDEgen}) we finally obtain 
 \begin{eqnarray}\label{wDEgenfin}
 &&
 \!\!\!\!\!\!\!\!\!\!\!\!\!\!
 w_{DE}=-1+\Omega_{DE}^{-1}\left\{ 
 \alpha\left[3+\frac{2\Omega_{DE}'}{1-\Omega_{DE}}\right.\right.
 \nonumber\\
 &&
\left. \left. \ \ \ \ \ \ \ \ \ \ \ \ \ \ \ \ \ \ \ \ \ \ \ 
 - 
\frac{2(\Omega_{DE}')^2}{(1-\Omega_{DE})^2}-\frac{\Omega_{DE}''}{1-\Omega_{DE}}\right]
\right.\nonumber\\
&&  \ \ \ \ \ \ \ \
- \frac{\beta}{\sqrt{3}(1-\Omega_{DE})^2}
\frac{1}{\sqrt{\left|\frac{\Omega_{DE}'}{1-\Omega_{DE}}-1
\right|}}\nonumber\\
&& \ \ \ \ \ \ \ \
\cdot\Big\{2\left[\Omega_{
DE}'-3(1-\Omega_{DE})
\right](\Omega_{DE}'+\Omega_{DE}-1)\nonumber\\
&&
 \ \ \ \ \ \ \ \ \left. \ \ 
+\Omega_{DE}''(1-\Omega_{DE})+(\Omega_{DE}
')^2
\Big\}
  \right\}.
\end{eqnarray}
In summary, since $\Omega_{DE}$ as a function of $\ln a$ is known from   (\ref{gensol}), 
 relation (\ref{wDEgenfin}) provides the equation-of-state parameter for 
Ricci-Gauss-Bonnet holographic dark energy as a function of $\ln a$, i.e as a function of 
the redshift. Lastly, it proves convenient to introduce the deceleration parameter $q$, 
given by 
  \begin{eqnarray}
  \label{qdeccel}
  q=-1-\frac{\dot{H}}{H^2}=\frac{1}{2}+\frac{3}{2}\left(w_m\Omega_m+w_{DE}\Omega_{DE}
  \right).
\end{eqnarray}

 \begin{figure}[ht]
\includegraphics[scale=0.45]{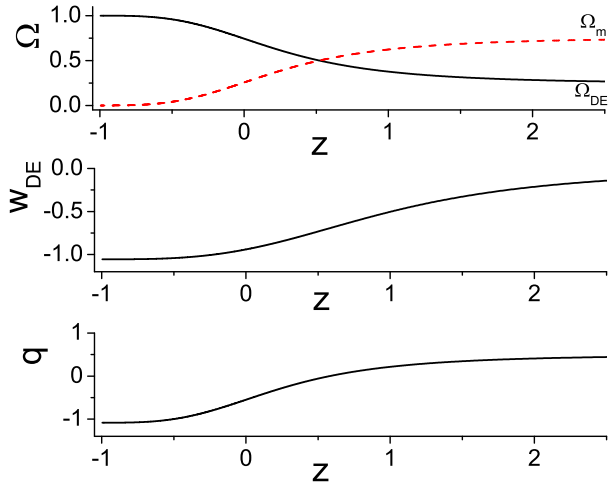}
\caption{
{\it{ Upper graph: The evolution of the Ricci-Gauss-Bonnet holographic dark energy    
density parameter $\Omega_{DE}$ (black-solid)  and of the matter density
parameter $\Omega_{m}$ (red-dashed), as a function of the redshift $z$,
for $\alpha=0.08$ and $\beta=0.001$, in units where $\kappa^2=1$. We have fixed 
 the integration constant  $x_0$ in (\ref{gensol}) in order to obtain at present 
$\Omega_{DE}(z=0)\equiv\Omega_{DE0}\approx0.68$.
  Middle graph: The evolution of the corresponding dark
energy equation-of-state parameter $w_{DE}$. Lower graph:  The evolution of the 
corresponding   deceleration parameter $q$.
}} }
\label{Omw1}
\end{figure}
Let us now use the above expressions in order to examine the evolution of the 
Ricci-Gauss-Bonnet holographic dark energy density $\Omega_{DE}$ and equation-of-state 
$w_{DE}$ parameters in terms of the redshift $z$, given straightforwardly through 
$x\equiv\ln a=-\ln(1+z)$. In the upper graph of Fig. \ref{Omw1} we present 
$\Omega_{DE}(z)$ and $\Omega_{m}(z)=1-\Omega_{DE}(z)$, as they are  given by the positive 
branch of  (\ref{gensol}). In the middle graph we depict the 
corresponding behavior of $w_{DE}(z)$ as it arises from  (\ref{wDEgenfin}). And in the 
lower graph we draw the deceleration parameter from (\ref{qdeccel}). As we 
mentioned above, the integration constant $x_0$ in (\ref{gensol}) has been chosen in order 
to obtain $\Omega_{DE}(z=0)\equiv\Omega_{DE0}\approx0.68$ in agreement with observations 
\cite{Ade:2015xua}. Finally, in the figures we have extended the evolution up to the far 
future, namely up to $z\rightarrow-1$ which corresponds to $t\rightarrow\infty$. 
As we 
observe, we can obtain the usual thermal history of the universe, with the transition 
from deceleration to acceleration happening at $z\approx 0.45$ as required, and in the 
future the universe tends asymptotically to a complete dark-energy dominated state.

\begin{figure}[ht]
\includegraphics[scale=0.45]{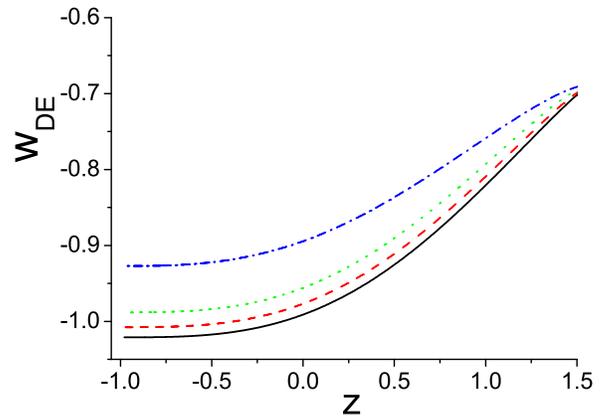}
\caption{
{\it{ The evolution of the equation-of-state parameter  $w_{DE}$  of the 
Ricci-Gauss-Bonnet 
holographic dark energy, as a function of the redshift $z$, for   $\alpha=0.08$ and 
$\beta=0.005$ (black-solid),     
$\alpha=0.08$ and $\beta=0.007$ (red-dashed), $\alpha=0.08$ and $\beta=0.01$ 
(green-dotted), and     $\alpha=0.08$ and $\beta=0.02$ (blue-dashed-dotted),  in units 
where $\kappa^2=1$.
We have fixed 
 the integration constant  $x_0$ in (\ref{gensol}) in order to obtain at present 
$\Omega_{DE}(z=0)\equiv\Omega_{DE0}\approx0.68$.
}} }
\label{wplo1}
\end{figure}

\begin{figure}[ht]
\includegraphics[scale=0.45]{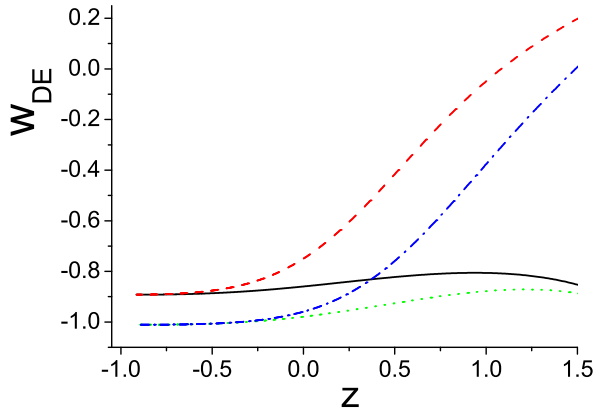}
\caption{
{\it{ The evolution of the equation-of-state parameter $w_{DE}$ of the Ricci-Gauss-Bonnet 
holographic dark energy, as a function of the redshift $z$, for   $\alpha=0.05$ and 
$\beta=0.1$ (black-solid),     
$\alpha=0.05$ and $\beta=-0.1$ (red-dashed), $\alpha=0.05$ and $\beta=0.08$ 
(green-dotted), and     $\alpha=0.05$ and $\beta=-0.08$ (blue-dashed-dotted),  in units 
where $\kappa^2=1$.
We have fixed 
 the integration constant  $x_0$ in (\ref{gensol}) in order to obtain at present 
$\Omega_{DE}(z=0)\equiv\Omega_{DE0}\approx0.68$.}} }
\label{wplo2}
\end{figure}

We mention that according to (\ref{wDEgen}), even if $\Omega_{DE}\rightarrow1$ at 
$z\rightarrow-1$, the asymptotic value of $w_{DE}$ depends on the model parameters 
$\alpha$ and $\beta$. In order to observe this behavior more transparently, in Figs. 
\ref{wplo1} and  \ref{wplo2} we present $w_{DE}(z)$ for various choices of $\alpha$ and 
$\beta$. In all cases the parameter  $x_0$ in 
(\ref{gensol}) has been chosen in order to obtain 
$\Omega_{DE}(z=0)\equiv\Omega_{DE0}\approx0.68$, and additionally the specific pairs of 
$\alpha$ and 
$\beta$ have been chosen in order to obtain an early-time behavior of $\Omega_{DE}(z)$ in 
agreement with observations (i.e. similarly  to Fig. \ref{Omw1}).

As we can see in Fig.  \ref{wplo1}, keeping $\alpha$ constant for increasing $\beta$  we 
obtain an increasing current (i.e. at $z=0$) and asymptotic future (i.e. at 
$z\rightarrow-1$) value of  $w_{DE}$. Thus, for some parameter regions the dark energy 
can exhibit the phantom-divide crossing before or after the present time, for some 
other parameter regions it lies always in the quintessence regime, and for some other
parameter regions  $w_{DE}$ can asymptotically go to $-1$ i.e. the universe results in a 
de Sitter phase (one can see that Eq. (\ref{FR1}) admits a late-time de Sitter 
solution if $1=12 \alpha+2\sqrt{3}\beta$). For completeness we 
mention that one can have parameter regions in which   $w_{DE}$ lies always in the 
phantom regime, and in some such cases the universe results to a Big Rip. 
Additionally, in  Fig.  \ref{wplo2} we can see that although the 
asymptotic value of  $w_{DE}$ does not depend on the sign of $\beta$ (one can easily see 
by taking the approximation of (\ref{gensol}) for $\Omega_{DE}\rightarrow1$  and then of 
(\ref{wDEgenfin}) that the resulting $w_{DE}$ depends only on $\beta^2$), its evolution 
depends significantly on that.  Definitely, in order to thoroughly investigate the 
future behavior of the scenario at hand at late times, and extract the asymptotic 
solutions and their stability, one should apply   dynamical system methods 
\cite{Copeland:1997et,Bahamonde:2017ize}, however such a detailed analysis lies beyond the 
scope of this work.

We note here that in Figs.  \ref{wplo1} and \ref{wplo2} we desire to present  
the effects of the parameters $\alpha$ and $\beta$ on the cosmological behavior. If one 
wishes additionally to  obtain $|w_{DE} + 1|\lesssim 0.1$ over the redshift interval $z = 
[0, 1.5)$ as it is favored by observations, then he should consider only 
e.g. the black-solid and green-dotted curves of \ref{wplo2}. Definitely,  the complete 
confrontation with 
the observed behavior can only be obtained using detailed data from Type Ia Supernovae (SN 
Ia), Baryon Acoustic Oscillations (BAO), Cosmic Microwave Background (CMB) shift 
parameter, and Hubble parameter observations, and perform the appropriate analysis.

At this point we should mention that for $\beta=0$, i.e. for the case of usual Ricci 
dark energy, the above 
expressions are significantly simplified. In particular, (\ref{gensol}) can be inverted 
giving  $\Omega_{DE}$ in an explicit form as
 \begin{eqnarray}
  \Omega_{DE}|_{Ric}=\frac{e^{\frac{x}{3\alpha}+c_1}-3\alpha   e^{  x  + 3\alpha c_1 }}{
  e^{\frac{x}{3\alpha}+c_1}-   e^{  x  + 3\alpha c_1    }},
 \end{eqnarray}
and then  (\ref{Hrel}) leads to 
 \begin{eqnarray}
H|_{Ric}= H_0\sqrt{\Omega_{m0}} e^{\frac{-3x}{2}}
\left[ \frac{(3\alpha-1)  e^{  x  + 3\alpha c_1    } }{ e^{\frac{x}{3\alpha}+c_1}    -   
e^{  x  + 3\alpha c_1    }}
\right]^{-1/2},
 \end{eqnarray}
and therefore inserting these into (\ref{OmDE}) gives
  \begin{equation}
 \rho_{DE}|_{Ric}=\frac{3H_0^2\Omega_{m0}}{\kappa^2(3\alpha-1)}
 \left[e^{-c_1(3\alpha-1)}
 e^{\left(\frac{1}{3\alpha}-4
 \right)x}   -3\alpha e^{-3x}
 \right],
 \end{equation}
which coincides exactly  with the result of \cite{Gao:2007ep} up to the re-definition of 
constants (the authors of that manuscript use a coupling $\alpha$ which is 6 times our 
$\alpha$ and they set $\kappa^2$ to $8\pi$). Finally, in this case (\ref{wDEgenfin}), 
using also (\ref{OmDEdifeq}), is simplified to 
   \begin{eqnarray}
w_{DE}|_{Ric}= \frac{1}{3\Omega_{DE}}-\frac{1}{\alpha}.
 \end{eqnarray}

On the other hand, for $\alpha=0$ we obtain a pure Gauss-Bonnet holographic dark energy. 
In this case the solution is obtained from  (\ref{gensol}) setting  $\alpha=0$, however 
although significantly simpler the resulting expression cannot be inverted. Nevertheless, 
 (\ref{wDEgenfin}), 
using also (\ref{OmDEdifeq}) can be greatly simplified, giving
   \begin{eqnarray}
w_{DE}|_{GB}= -\frac{1}{3\Omega_{DE} }  \left(\frac{\Omega_{DE}^2}{12\beta^2}+1\right).
 \end{eqnarray}

 We close this section by investigating the above scenario under two extensions. The 
first is when we allow for an interaction between the Ricci-Gauss-Bonnet holographic 
dark energy and the dark-matter sector. The second is when we include also the radiation 
sector, in order to describe the whole thermal history of the universe.
 
 \begin{itemize}
 
 \item {Interacting case}
 
 In  a general dark-energy scenario one may allow for an interaction between dark-energy 
and dark-matter sectors, since on   one hand such an interaction cannot be excluded 
by 
theoretical arguments, and on the other hand it proves to lead to an alleviation of the 
so-called coincidence problem, namely why the density parameters of dark energy and dark 
matter are almost equal today although these sectors follow a completely different 
scaling during cosmological evolution \cite{Zimdahl:2001ar,Chen:2008ft}. 

In order to quantify the interaction we follow the standard procedure and we modify 
the conservation equations (\ref{rhocons}) and (\ref{rhoDEcons}) as
\begin{eqnarray}\label{rhocons22}
&&\dot{\rho}_m+3H(\rho_m+p_m)=-Q\\
&&\label{rhoDEcons22}
\dot{\rho}_{DE}+3H(1+w_{DE})\rho_{DE}=Q,
\end{eqnarray}
where $Q$ is a phenomenological descriptor of the 
interaction. 
Thus, $Q>0$ corresponds to energy transfer from the dark-matter sector to the
dark-energy one, whereas $Q<0$ corresponds to transformation of dark energy 
to dark matter. Although the form of $Q$ can be chosen at will,  a well-studied choice 
is to assume that $Q=\xi H \rho_m$, with $\xi$ a parameter, which has a reasonable 
justification since the interaction rate is indeed expected to be proportional to the 
energy 
density \cite{Zimdahl:2001ar,Chen:2008ft}.
 
Let us now see how the Ricci-Gauss-Bonnet holographic dark energy cosmology will change 
in 
the presence of the above interaction term $Q$. Firstly, according to (\ref{rhocons22}) 
the matter evolution will become
    \begin{eqnarray}
\rho_m=\frac{\rho_{m0}}{a^{3+\xi}},
 \end{eqnarray}
 and therefore (\ref{Hrel}) will become
  \begin{equation}\label{Hrel22}
H=\frac{H_0\sqrt{\Omega_{m0}}}{\sqrt{a^{3+\xi}(1-\Omega_{DE})}}.
\end{equation}
Hence,  (\ref{Hdorrel}) will extend to
 \begin{equation}\label{Hdorrel22}
\dot{H}=-\frac{H^2}{2(1-\Omega_{DE})}\left[(3+\xi)(1-\Omega_{DE})-\Omega_{DE}'
\right],
\end{equation}
and then (\ref{RicH2}), (\ref{GBH2}) will read as
  \begin{eqnarray}
 && R=-3 H^2 \left[1-\xi+\frac{\Omega_{DE}'}{1-\Omega_{DE}}
 \right]
 \label{RicH222}\\
 &&G=12H^4 \left[\frac{\Omega_{DE}'}{1-\Omega_{DE}}-1-\xi
 \right]
  \label{GBH222} .
 \end{eqnarray}
 Thus, the differential equation (\ref{OmDEdifeq}) for $\Omega_{DE}$ extends to
  {\small{ \begin{equation}
  \Omega_{DE}=3\alpha \left[1\!-\!\xi\!+\!\frac{\Omega_{DE}'}{1\!-\!\Omega_{DE}}
 \right]+2\sqrt{3}\beta \sqrt{\left|\frac{\Omega_{DE}'}{1\!-\!\Omega_{DE}}\!-\!1-\!\xi
 \right|}
  \label{OmDEdifeq22} .
 \end{equation}}}
 
 The analytical solution of the above differential equation, which generalizes 
(\ref{gensol}) 
in the interacting case, writes as
{\small{   \begin{eqnarray}
 &&\!\!\!\!\!\!\!\!\!\!\!
 \epsilon_{\pm}  \frac{\gamma_+}{\delta \zeta_+\sqrt{1+\xi}}\, \text{arctanh} 
\left(-\frac{\sqrt{6\alpha^2+\beta^2-\alpha\,\Omega_{DE}}}{ \zeta_+}
 \right)\nonumber\\
 &&\!\!\!\!\!\!\!\!\!\!\!\! -  \epsilon_{\pm}   \frac{\gamma_-}{\delta 
\zeta_-\sqrt{1+\xi}}\, 
\arctan 
\left(-\frac{\sqrt{6\alpha^2+\beta^2-\alpha\,\Omega_{DE}}}{ \zeta_-}
 \right)\nonumber\\
  &&\!\!\!\!\!\!\!\!\!\!\!\!
 - 24 \epsilon_{\pm} 
\frac{\beta}{\delta}\sqrt{6\alpha^2+\beta^2-\alpha}    
 \, \text{arctanh} 
\left(\!\frac{\sqrt{6\alpha^2\!+\!\beta^2\!-\!\alpha\,\Omega_{DE}} 
}{\sqrt{6\alpha^2\!+\!\beta^2\!-\!\alpha}}
 \right)
 \nonumber\\
 &&\!\!\!\!\!\!\!\!\!\!\!\!\!
-\frac{\sqrt{3}[3\alpha^2(1-\xi)\!+\!2\beta^2\!-\!\alpha]}{\delta}\nonumber\\
&&\cdot
\ln\left\{
\frac {
(1+\xi)[
9\alpha^2(1+\xi)\!-\!12\beta^2\!-\!6\alpha\,
\Omega_{DE}]\! +\!\Omega_{DE}^2} {(\Omega_{DE}-1)^2 }
 \right\}
 \nonumber\\ 
 &&\!\!\!\!\!\!\!\!\!\!\!\!\!\!
 -\frac{2 [3\alpha(3\!+\!\xi)\!-\!1]\beta}{\delta\sqrt{1\!+\!\xi}}
 \, \text{arctanh} \!
\left[\frac{\Omega_{DE}\!-\!3\alpha(1\!-\!\xi)}{2\sqrt{3}\beta\sqrt{1\!+\!\xi}}
 \right]\!=\!2\ln a\!+\!x_0,\ \
 \label{gensol22}
 \end{eqnarray}}}
where $\epsilon_{\pm}=\pm1$, but now with
 \begin{eqnarray}
 && \!\!\!
 \gamma_{\mp}=2\left\{\mp  9 \alpha^3(1-\xi^2)
 +
\beta^2\left(\pm1+2\sqrt{3}\beta\sqrt{1+\xi}\right)
\right.\nonumber\\
&& \ \ \ \ \ \ \ \ \ \ 
\left. \mp\alpha \beta \left[\pm 
2\sqrt{3}\sqrt{1+\xi}+3\beta(5+3\xi)\right]
\right.
\nonumber\\
&& \ \ \ \ \ \ \ \ \ \ 
\left.
+3\alpha^2 
\left[4\sqrt{3}\beta\sqrt{1+\xi}\pm(1+\xi)\right]
 \right\},\nonumber\\ 
&&\delta=\frac{1-6\alpha(1-\xi)+9\alpha^2(1-\xi)^2-12\beta^2(1+\xi)}{\sqrt{3}},\nonumber\\
 &&\zeta_{\mp}=\sqrt{\mp3 \alpha^2(1+\xi)+2\sqrt{3}\alpha\beta\sqrt{1+\xi}\mp\beta^2}.
 \end{eqnarray}
 
 Hence, as discussed above,  $\xi>0$ corresponds to energy transfer from dark matter 
  to dark energy, in which case the dark-energy domination will take place sooner, 
while $\xi<0$ corresponds to transformation of dark energy 
to dark matter, in which case the dark-energy epoch will arise later.

   \item {Radiation inclusion}

As can be seen from the upper graph of Fig. 
\ref{Omw1}, the matter energy density tends to 1 as $z$ grows, and one can easily 
verify that this behavior is maintained for the whole $z$-interval that is required for 
the correct description of the matter era (namely up to $z\sim 2000$). Nevertheless, 
since we have not explicitly included radiation, this matter-era will be maintained for 
larger $z$ too, and thus in order to stop it and obtain a radiation era in agreement with 
the thermal history of the universe, we must include radiation. In the following we 
provide the involved equations of the scenario of Ricci-Gauss-Bonnet holographic 
dark energy in the presence of the radiation sector, since this will allow us to describe 
the whole thermal history of the universe. 

In particular,
considering 
\begin{eqnarray}
  \Omega_r\equiv\frac{\kappa^2}{3H^2}\rho_r,
 \label{OmR2}
 \end{eqnarray} 
 the Friedmann equation (\ref{FR1}) becomes 
  $\Omega_r+ \Omega_m+\Omega_{DE}=1$,  which leads to
 \begin{equation}
 \label{Hrel2}
H=\frac{H_0\sqrt{\Omega_{m0}}}{\sqrt{a^3(1-\Omega_r- \Omega_{DE})}}.
\end{equation}
Then (\ref{Hdorrel}) extends to 
{\small{
  \begin{equation}\label{Hdorrel2}
\dot{H}=-\frac{H^2}{2(1\!-\!\Omega_r\!-\! 
\Omega_{DE})}\Bigl[3(1-\Omega_r-\Omega_{DE})\!-\!\Omega_r'\!-\!\Omega_{DE}
'
\Bigr],
\end{equation}}}
and similarly (\ref{RicH2}) and (\ref{GBH2}) become 
  \begin{eqnarray}
 && R=-3 H^2 \left[1+\frac{\Omega_r ' + \Omega_{DE}'}{1-\Omega_r - \Omega_{DE}}
 \right]
 \label{RicH22}\\
 &&G=12H^4 \left[\frac{\Omega_r ' +\Omega_{DE}'}{1-\Omega_r - \Omega_{DE}}-1
 \right]
  \label{GBH22} .
 \end{eqnarray}
Inserting these into (\ref{rhoHDE}) and then into (\ref{OmDE}) we   
obtain 
   \begin{eqnarray}
 &&  \Omega_{DE}=3\alpha \left[1+\frac{\Omega_r '+ \Omega_{DE}'}{1-\Omega_r - \Omega_{DE}}
 \right]\nonumber\\
 &&
 \  \  \  \  \  \  \  \  \  \,
 +2\sqrt{3}\beta \sqrt{\left|\frac{\Omega_r '+ \Omega_{DE}'}{1-\Omega_r 
'- \Omega_{DE}}-1
 \right|}
  \label{OmDEdifeq2} .
 \end{eqnarray}
 Finally, (\ref{Ricprim}) and (\ref{GBprim}) are now extended to
 \begin{eqnarray}
 &&  \frac{R'}{3H^2}=3+2 \left(\frac{\Omega_r ' + \Omega_{DE}'}{1-\Omega_r - \Omega_{DE}} 
\right)- 
\frac{2(\Omega_r '+ \Omega_{DE}')^2}{(1-\Omega_r - 
\Omega_{DE})^2}\nonumber\\
&& \ \ \ \ \ \ \  \ \ \
-\frac{\Omega_r ''+ \Omega_{DE}''}
{1-\Omega_r - \Omega_{DE}}
\label{Ricprim2}
 \end{eqnarray}
and 
 \begin{eqnarray}
 &&   \frac{ 
|G'|}{6H^2\sqrt{|G|}}=\frac{1}{\sqrt{3}(1-\Omega_r-\Omega_{DE})^2}
\frac{1}{\sqrt{\left|\frac{\Omega_r'+\Omega_{DE}'}{1-\Omega_r-\Omega_{DE}}-1
\right|}}\nonumber\\
&& \ \ \ \ \ \ \  \ \ \ \ \ \ \  \ \ \
\cdot\Big\{\Omega_{DE}''(1-\Omega_r-\Omega_{DE})+(\Omega_r'+\Omega_{DE}
')^2\nonumber\\
&&
\ \ \ \ \ \ \  \ \ \ \ \ \ \  \ \ \ \ \ \,
-2\left [ 3(1-\Omega_r-\Omega_{DE} )-\Omega_r'-\Omega_{
DE}'
\right]\cdot\nonumber\\
&&\ \ \ \ \ \ \  \ \ \ \ \ \ \  \ \ \ \ \ \ \ 
(\Omega_r'+\Omega_{DE}'+\Omega_r+\Omega_{DE}-1)
\Big\},
\label{GBprim2}
 \end{eqnarray}
 while (\ref{wDEgenfin}) becomes
 \begin{eqnarray}\label{wDEgenfin2}
 &&w_{DE}=-1+\Omega_{DE}^{-1}
 \left\{ 
 \alpha\left[
 3+2 \left(\frac{\Omega_r ' + \Omega_{DE}'}{1-\Omega_r - \Omega_{DE}} 
\right) \right.\right.\nonumber\\
&&\left. \left. \ \ \ \ \ \ \  \ \ \ \ \ \ \ \ \  \ \ \ \ \ \ \ \ \ \ \ \ \  \
-
\frac{2(\Omega_r '+ \Omega_{DE}')^2}{(1-\Omega_r - 
\Omega_{DE})^2}\right.\right.\nonumber\\
&& \left.  \ \ \ \ \ \ \  \ \ \ \ \ \ \ \ \  \ \ \ \ \ \  \ \ \ \ \ \ \ \ \,
-\frac{\Omega_r ''+ \Omega_{DE}''}
{1-\Omega_r - \Omega_{DE}}
 \right]
 \nonumber\\
&&
- \frac{\beta}{\sqrt{3}(1-\Omega_r-\Omega_{DE})^2}
\frac{1}{\sqrt{\left|\frac{\Omega_r'+\Omega_{DE}'}{1-\Omega_r-\Omega_{DE}}-1
\right|}}\nonumber\\
&& \ \ \ \ \ \ \  \ \ \ 
\cdot
\Big\{\Omega_{DE}''(1-\Omega_r-\Omega_{DE})+(\Omega_r'+\Omega_{DE}
')^2\nonumber\\
&& 
\ \ \ \ \ \ \  \ \ \  \  \ \,
-2\left [ 3(1-\Omega_r-\Omega_{DE} )-\Omega_r'-\Omega_{
DE}'
\right]\cdot\nonumber\\
&&
\left.
\ \ \ \ \ \ \  \ \  \   \ \ \ \ 
(\Omega_r'+\Omega_{DE}'+\Omega_r+\Omega_{DE}-1)
\Big\}
  \right\}.
\end{eqnarray}
Contrary to (\ref{OmDEdifeq}), differential equation (\ref{OmDEdifeq2}) does not accept 
an analytical solution, and thus one should elaborate it numerically.

In Fig. \ref{Omradiation} we present the evolution of the Ricci-Gauss-Bonnet holographic 
dark energy density parameter $\Omega_{DE}$, of the matter density
parameter $\Omega_{m}$, and of the radiation density
parameter $\Omega_{r}$,  in the case where 
radiation is included in the scenario,  imposing 
$\Omega_{DE0}\approx0.68$, 
$\Omega_{m0}\approx0.32$ and 
$\Omega_{r0}\approx0.0001$ at present time, as required by observations. As we observe, 
we can obtain the thermal history of the universe, with the successive sequence of 
radiation, matter, and dark energy eras, before the universe result in complete 
dark-energy domination.
 \begin{figure}[ht]
\includegraphics[scale=0.45]{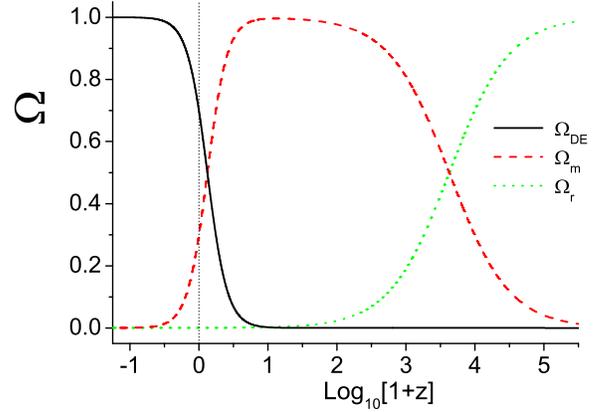}
\caption{
{\it{The evolution of the Ricci-Gauss-Bonnet holographic dark energy    
density parameter $\Omega_{DE}$ (black-solid), of the matter density
parameter $\Omega_{m}$ (red-dashed), and of the radiation density
parameter $\Omega_{r}$ (green-dotted), as a function of the redshift $z$,
for $\alpha=0.05$ and $\beta=0.12$ in units where $\kappa^2=1$, in the case where 
radiation is included in the scenario. We have imposed 
$\Omega_{DE}(z=0)\equiv\Omega_{DE0}\approx0.68$, 
$\Omega_{m}(z=0)\equiv\Omega_{m0}\approx0.32$, and 
$\Omega_{r}(z=0)\equiv\Omega_{r0}\approx0.0001$ at present time, marked by the dotted 
vertical line.  
}} }
\label{Omradiation}
\end{figure}
 
 \end{itemize}

\section{ Constraints from Big Bang Nucleosynthesis}

As one can see from expression (\ref{rhoHDE2}), Ricci-Gauss-Bonnet holographic dark 
energy is not necessarily zero at early times. Hence, apart from the basic 
phenomenological requirements imposed in the previous section, one should carefully 
examine the constraints that should be imposed on the scenario from Big Bang 
Nucleosynthesis (BBN). Hence, in this Section we examine the  BBN constraints in detail.

The energy density of relativistic particles constituting the Universe  is
  ${\displaystyle \rho_r=\frac{\pi^2}{30}g_* {{T}}^4}$,
where $g_*\sim 10$ is
the effective number of degrees of freedom and ${{T}}$ the temperature (in the
Appendix we review the main
features related to the BBN physics). The neutron abundance is computed via the
conversion rate of protons into neutrons, namely
 \[
 \lambda_{pn}({{T}})=\lambda_{n+\nu_e\to p+e^-}+\lambda_{n+e^+\to p+{\bar
\nu}_e}+\lambda_{n\to p+e^- +
{\bar \nu}_e}\,,
\]
and its inverse $\lambda_{np}({{T}})$, and the relevant quantity is the total rate
 \begin{equation}\label{Lambda}
    \Lambda({{T}})=\lambda_{np}({{T}})+\lambda_{pn}({{T}})\,.
 \end{equation}
Explicit calculations of Eq. (\ref{Lambda})   (see (\ref{LambdafinApp}) in the
Appendix) give
 \begin{equation}\label{Lambdafin}
    \Lambda({{T}}) =4 A\, {{T}}^3(4! {{T}}^2+2\times 3! {\cal Q}{{T}}+2!
{\cal Q}^2)\,,
 \end{equation}
where ${\cal Q}=m_n-m_p$ is the mass difference of neutron and proton, and $A=1.02 \times
10^{-11}$GeV$^{-4}$. The primordial mass
fraction of ${}
^4 He$ can be estimated by making use of the relation \cite{kolb}
 \begin{equation}\label{Yp}
    Y_p\equiv \lambda \, \frac{2 x(t_f)}{1+x(t_f)}\,,
 \end{equation}
where $\lambda=e^{-(t_n-t_f)/\tau}$, with $t_f$   the time of the freeze-out of the weak
interactions, $t_n$   the time of the freeze-out of the nucleosynthesis,
$\tau$ the neutron mean lifetime given in (\ref{rateproc3}), and
$x(t_f)=e^{-{\cal
Q}/{{T}}(t_f)}$   the neutron-to-proton equilibrium ratio.
The function $\lambda(t_f)$ is interpreted as the fraction of neutrons that decay into
protons during the interval $t\in [t_f, t_n]$. Deviations from the fractional mass $Y_p$
due to
the variation of the freezing temperature ${{T}}_f$ are given by
 \begin{equation}\label{deltaYp}
    \delta
Y_p=Y_p\left[\left(1-\frac{Y_p}{2\lambda}\right)\ln\left(\frac{2\lambda}{Y_p}
-1\right)-\frac{2t_f}{\tau}\right]
    \frac{\delta {{T}}_f}{{{T}}_f}\,,
 \end{equation}
where we have set $\delta {{T}}(t_n)=0$ since ${{T}}_n$ is fixed by the deuterium
binding energy
\cite{torres,Lambiase1}. The experimental estimations of the mass
fraction $Y_p$ of baryon
converted to ${}^4 He$
during the Big Bang Nucleosynthesis are
$
 Y_p=0.2476$ and $ |\delta Y_p| < 10^{-4}\,
$ \cite{coc,altriBBN1}.
 Inserting these into (\ref{deltaYp}) one infers the upper bound on $\frac{\delta
{{T}}_f}{{{T}}_f}$, namely
 \begin{equation}
 \label{deltaT/Tbound}
    \left|\frac{\delta {{T}}_f}{{{T}}_f}\right| < 4.7 \times 10^{-4}\,.
 \end{equation}

During the BBN, at the radiation dominated era, the scale factor evolves as $a\sim
t^{1/2}$, where $t$ is cosmic time. The  Ricci-Gauss-Bonnet holographic dark energy  
  density $\rho_{DE}$ is
treated as a perturbation to the radiation energy density $\rho_r$. The relation between
the
cosmic time and the temperature
is given by ${\displaystyle \frac{1}{t}\simeq \left(\frac{32\pi^3 
g_*}{90}\right)^{1/2}\frac{{
T}^2}{M_{P}}
}$
(or
${{T}}(t)\simeq (t/\text{sec})^{1/2} $MeV). Furthermore, we use the entropy
conservation
$S\sim a^3
{{T}}^3=constant$. The expansion rate of
the Universe  is derived from (\ref{FR1}), and can be rewritten in the form
  \begin{eqnarray}\label{H+H1}
    H&=&H_{GR}^{(R)}\sqrt{1+\frac{\rho_{DE}}{\rho_r}}=H_{GR}+\delta H\,,  \\
    \delta H&=&\left(\sqrt{1+\frac{\rho_{DE}}{\rho_r}}-1\right)H_{GR}\,,
 \end{eqnarray}
where $H_{GR}=\displaystyle{\sqrt{\frac{\kappa^2}{2}\rho_r}}$ ($H_{GR}$ is the expansion 
rate
of the
Universe  in general relativity). Thus, from the relation $\Lambda= H$, one derives the
freeze-out
temperature
${{T}}={{T}}_f\left(1+\frac{\delta {{T}}_f}{{{T}}_f}\right)$, with ${T}_f\sim 0.6$ MeV 
(which follows from
$H_{GR}\simeq q {{T}}^5$)
and
\begin{equation}\label{H_T=Lambda}
  \left(\sqrt{1+\frac{\rho_{DE}}{\rho_r}}-1\right)H_{GR} = 5q {{T}}_f^4 \delta {{T}}_f\,,
  \end{equation}
from which, in the regime $\rho_{DE}\ll \rho_r$, one obtains:
  \begin{equation}\label{deltaT/TboundG}
  \frac{\delta {{T}}_f}{{{T}}_f}\simeq \frac{\rho_{DE}}{\rho_r}\frac{H_{GR}}{10 q
{{T}}_f^5}\,,
\end{equation}
with $q=4! A\simeq 9.6\times 10^{-36}$GeV$^{-4}$.

Let us now investigate the bounds that arise from the BBN constraints, on
the free parameters $\alpha$ and $\beta$ of the scenario of Ricci-Gauss-Bonnet 
holographic dark energy. 
These constraint will be determined using Eqs. (\ref{deltaT/TboundG}) and 
(\ref{rhoHDE2}).  Assembling everything (we neglect the last two terms in the 
parenthesis of (\ref{LambdafinApp})) we find
  \begin{equation}\label{newrell}
  \frac{\delta {{T}}_f}{{{T}}_f}\simeq
  \frac{16q^2}{\pi^5}\left(\frac{90}{\kappa^2 g_*}
  \right)^{5/2}\left(\frac{T_f}{T_0}\right)^8 H_0^2 \Omega_{m0}^2 
\left(12\alpha+2\sqrt{3}\beta\right),
  \end{equation}
with $ {{T}}_0$, $H_0$ and $\Omega_{m0}$ the current values of the  CMB temperature, of 
the Hubble parameter and of the matter density parameter, respectively.  Hence, using  
the bound (\ref{deltaT/Tbound}) we finally find that 
  \begin{equation}\label{limitsss}
  12\alpha+2\sqrt{3}\beta < 2.53,
  \end{equation}
  in units where $\kappa^2=1$. As we can see, the examples considered in the previous 
section, where prove to exhibit a cosmological behavior in agreement with observations, 
as well as the de Sitter solution condition, lie within the above bound.

\section{Conclusions}
\label{Conclusions}

In this work we presented a model of holographic dark energy in which the Infrared cutoff 
is determined by both the Ricci scalar and the Gauss-Bonnet invariant. Such a 
construction has the significant advantage that the Infrared cutoff, and consequently the 
holographic dark energy density, does not depend on the future (such as in standard 
holographic dark energy versions) or the past (such as in agegraphic dark energy 
versions) evolution of the universe, but only on its current features. Additionally, it 
has the theoretical advantage that the Infrared cutoff is determined by invariants, whose 
role is fundamental in gravitational theories. Finally, following the usual approach in 
such theories, we allowed for more than one invariants of the same order to contribute.
 
 In order to investigate the cosmological applications of the Ricci-Gauss-Bonnet 
holographic dark energy we first elaborated the equations and we
resulted to a simple differential equation for the holographic dark energy density 
parameter $\Omega_{DE}$ in terms of the logarithm of the scale factor (which is 
straightforwardly related to the logarithm of the redshift), that accepts an analytical 
solution. Furthermore, we extracted the form of the holographic dark energy 
equation-of-state parameter $w_{DE}$ as a function of $\Omega_{DE}$, and thus resulting  
to its form as a function of the redshift.

The scenario of  Ricci-Gauss-Bonnet holographic dark energy leads to interesting 
cosmological behavior, with enhanced capabilities due to the presence of two model 
parameters. First of all one can obtain the usual thermal history 
of the universe, with a dark-matter era followed by a dark-energy one, where the onset 
of acceleration takes place at $z\approx0.45$ in agreement with observations, resulting 
in the future to a complete dark energy domination. The corresponding dark energy 
equation-of-state parameter can have a rich behavior, lying in the quintessence regime, 
in the phantom regime, or experiencing the phantom-divide crossing during the 
cosmological 
evolution. Moreover, its asymptotic value in the far future is determined by the model 
parameters, and can be quintessence-like, phantom-like, or be exactly equal to the 
cosmological-constant value $-1$. In the simple case where the contribution 
of the  Gauss-Bonnet invariant is set to zero we re-obtained the results of usual Ricci 
dark energy, while in the case where  the contribution of the  Ricci invariant is set to 
zero, we obtained a ``pure'' Gauss-Bonnet holographic dark energy, where the various 
expressions are simplified.

In the case where radiation is included, one can describe the whole thermal  
 history of the universe, namely the successive sequence of 
radiation, matter, and dark energy eras, before the universe result in complete 
dark-energy domination. Finally, examining in detail the constraints that arise 
from Big Bang Nucleosynthesis, we showed that in the above solutions the chosen model 
parameters satisfy the obtained bound.

We mention here that the resulting form of the Ricci-Gauss-Bonnet holographic dark energy 
contains terms depending on the Hubble function and its derivative. This might be similar 
to some particular models of the ``running vacuum'' classes  
\cite{Basilakos:2012ra,George:2015lok,Sola:2016jky,Basilakos:2011wm}, nevertheless the 
significant advantage is that in the present scenario they are justified under the 
holographic considerations using invariants, while in the former classes of models they 
are introduced by hand. Similarly, the scenario at hand is a sub-class of the 
general covariant generalized holographic dark energy, in which the Infrared cutoff is an 
arbitrary function of the Hubble function, the particle and future horizons, the
cosmological constant, the universe age and their derivatives 
\cite{Nojiri:2017opc}, however in the present approach it is a theoretically 
justified and well-defined sub-class.

We close this work by mentioning that there are additional investigations that need to be 
done before  the scenario of  Ricci-Gauss-Bonnet holographic dark energy can be 
considered as a successful candidate for the description of nature. One should use data 
from Type Ia Supernovae (SN Ia), Baryon Acoustic Oscillations (BAO),
Cosmic Microwave Background (CMB) shift parameter, and Hubble parameter observations, 
in order to extract constraints on the parameters of the scenario. Additionally, he 
should perform a perturbation analysis and confront it with perturbation-related data 
such as CMB temperature and polarization, Large Scale Structure (LSS) and gravitational 
lenses. Furthermore, he should apply the dynamical system methods in order to reveal the 
global behavior of the scenario at late times, independently of the early- and 
intermediate- 
time evolutions. These necessary studies lie beyond the scope of this manuscript and are 
left for future  investigations.

\section*{Acknowledgments}
The author wishes to thank National Center for Theoretical Sciences,
Hsinchu, Taiwan for the hospitality during the preparation of
this work.

\appendix*

\section{Big Bang Nucleosynthesis}

 In this Appendix  we briefly review the main features of Big Bang Nucleosynthesis (BBN)
following
\cite{kolb,bernstein}. As it is known, in the early Universe the primordial ${}^4He$ was 
formed at temperature ${{T}}\sim 100$ MeV.   Protons and neutrons were 
maintained in 
thermal
equilibrium due to the interactions 
 \begin{eqnarray}\label{proc1}
    \nu_e+n & \longleftrightarrow & p+e^- \\
    e^++n & \longleftrightarrow & p + {\bar \nu}_e \label{proc2} \\
    n& \longleftrightarrow & p+e^- + {\bar \nu}_e\,. \label{proc3}
 \end{eqnarray}
The neutron abundance is estimated by computing the conversion rate of protons into
neutrons  $\lambda_{pn}({{T}})$, and its inverse $\lambda_{np}({{T}})$. Therefore,
the weak interaction rates at suitably high temperature 
read as
 \begin{equation}\label{LambdaA}
    \Lambda({{T}})=\lambda_{np}({{T}})+\lambda_{pn}({{T}})\,,
 \end{equation}
where the rate $\lambda_{np}$ is given by
 \begin{equation}\label{sumprocess}
    \lambda_{np}=\lambda_{n+\nu_e\to p+e^-}+\lambda_{n+e^+\to p+{\bar
\nu}_e}+\lambda_{n\to p+e^- +
{\bar \nu}_e}\,.
 \end{equation}
 Finally, the rate $\lambda_{np}$ is related to the rate $\lambda_{pn}$ as
$\lambda_{np}({{T}})=e^{-{\cal Q}/{{T}}}\lambda_{pn}({{T}})$, with ${\cal
Q}=m_n-m_p$ the mass difference of neutron and proton.

During the freeze-out stage, one can use the following approximations
\cite{bernstein}: (i) The
temperatures of particles are the same, i.e. ${{T}}_\nu={{T}}_e={T}_\gamma={{T}}$. (ii) 
The temperature
${{T}}$ is lower than the typical energies $E$ that contribute to the
integrals entering the definition of the rates (one can thus replace the
Fermi-Dirac distribution with the Boltzmann
one, namely $n\simeq e^{-E/{{T}}}$). (iii) The electron mass $m_e$ can be neglected
with
respect to the
electron and neutrino energies ($m_e\ll E_e, E_\nu$).

Having these in mind, the interaction rate
corresponding to the process (\ref{proc1}) is given by
  \begin{equation}\label{rateproc1}
    d\lambda_{n+\nu_e\to p+e^-}= d\mu \,(2\pi)^4 |\langle{\cal M}|^2\rangle W \,,
  \end{equation}
where
\begin{eqnarray}
  d\mu & \equiv &  \frac{d^3p_e}{(2\pi)^3 2E_e} \frac{d^3p_{\nu_e}}{(2\pi)^3
2E_{\nu_e}}\frac{d^3p_
p}{(2\pi)^3 2E_p}\,, \label{dmu} \\
  W &\equiv & \delta^{(4)}({\cal P})n(E_{\nu_e})[1-n(E_e)]\,, \label{WA}
   \end{eqnarray}
   \begin{eqnarray}
   {\cal P}  &   \equiv &  p_n+p_{\nu_e}-p_p-p_e\,,  \\
 {\cal M} &= &\left(\frac{g_w}{8M_W}\right)^2 [{\bar u}_p\Omega^\mu u_n][{\bar
u}_e\Sigma_\mu v_{\nu_e}]\,, \label{M} \\
  \Omega^\mu &\equiv & \gamma^\mu(c_V-c_A \gamma^5)\,,
  \\
  \Sigma^\mu&
  \equiv&
\gamma^\mu(1-\gamma^5)
\,.
 \end{eqnarray}
In (\ref{M}) we have used the condition $q^2 \ll M_W^2$, where $M_W$ is the mass of the
vector gauge boson $W$, with $q^\mu=p_n^\mu-p_p^\mu$  the transferred
momentum. From expression (\ref{rateproc1}) it follows that
$
    \lambda_{n+\nu_e\to p+e^-}=A \, {{T}}^5 I_y\,,
$
where
$
  A\equiv \frac{g_V+3g_A}{2\pi^3}\,,
$
and where
$
 I_y=\int_y^\infty \epsilon(\epsilon-{\cal Q}')^2\sqrt{\epsilon^2-y^2}\, n(\epsilon-{\cal
Q})[1-n(\epsilon)]d\epsilon,
$
 with
$
 y\equiv \frac{m_e}{{{T}}}\,, \quad {\cal Q}'=\frac{{\cal Q}}{{{T}}}$.

  A similar calculation for the process (\ref{proc2}) gives
$
    \lambda_{e^+ + n \to p+ {\bar \nu}_e}=A \, {{T}}^5 J_y\,,
$
 with
$
 J_y=\int_y^\infty \epsilon(\epsilon+{\cal Q}')^2\sqrt{\epsilon^2-y^2}\,
n(\epsilon)[1-n(\epsilon+{\cal Q}')]d\epsilon\,,
$
 which finally  results to
$
 \label{ne-pnu-fin}
    \lambda_{e^+ + n\to p+{\bar \nu}_e}=A\, {{T}}^3(4! {{T}}^2+2\times 3! {\cal
Q}{{T}}+2! {\cal
Q}^2)\,.
$

Lastly, for the neutron decay (\ref{proc3}) one obtains
 \begin{equation}\label{rateproc3}
    \tau=\lambda_{n\to p+e^- +{\bar \nu}_e}^{-1}\simeq 887 \text{sec}\,.
 \end{equation}
 Hence, in the calculation of (\ref{sumprocess}) we can safely neglect the above
interaction rate of the neutron decay, i.e. during the BBN the neutron can be
considered as a stable particle.

The above approximations (i)-(iii) lead to
\cite{bernstein}
 \begin{equation}
 \label{auxilirel}
  \lambda_{e^+ +n\to p+{\bar \nu}_e}=\lambda_{n+\nu_e\to p+e^-}\,.
 \end{equation}
 Thus, inserting   (\ref{auxilirel}) into (\ref{sumprocess}), and then
into (\ref{LambdaA}), allows to derive the expression for
$\Lambda({{T}})$, namely
 \begin{equation}\label{LambdafinA}
    \Lambda({{T}})\simeq 2\lambda_{np}=4\lambda_{e^+ +n\to p+{\bar \nu}_e}\,,
 \end{equation}
 which using the above expressions leads to
  \begin{equation}\label{LambdafinApp}
    \Lambda({{T}}) =4 A\, {{T}}^3(4! {{T}}^2+2\times 3! {\cal Q}{{T}}+2!
{\cal Q}^2)\,.
 \end{equation}

\end{document}